\documentclass[preprint]{emulateapj}
\usepackage{epsf}
\usepackage{apjfonts}
\usepackage{xspace}
\usepackage{amsmath}
\def\figdir{.}

\def\HII{\mbox{\ion{H}{2}}}

\def\H2{\mbox{H$_2$}\xspace}

\def\ps{P_{\rm S}}
\def\vm{V_{\rm max}}
\def\vtf{V_{\rm F}}
\def\vf{V_{\rm rf}}
\def\Lsun{\, L_{\odot}}
\def\Msun{\, M_{\odot}}
\def\chimp{h^{-1}\dim{Mpc}}
\def\kms{\dim{km/s}}
\def\MW{Milky Way}

\def\dim#1{\mbox{\,#1}}
\def\figname#1{\figdir/#1}

\def\hide#1{}

\begin{document}

\title{Fossils of Reionization in the Local Group}

\author{Nickolay Y.\
  Gnedin\altaffilmark{1,3} and Andrey V. Kravtsov\altaffilmark{2,3}}  
\altaffiltext{1}{Particle Astrophysics Center, 
Fermi National Accelerator Laboratory, Batavia, IL 60510, USA; gnedin@fnal.gov}
\altaffiltext{2}{Kavli Institute for Cosmological Physics and Enrico
  Fermi Institute, The University of Chicago, Chicago, IL 60637 USA;
  andrey@oddjob.uchicago.edu} 
\altaffiltext{3}{Department of Astronomy \& Astrophysics, The
  University of Chicago, Chicago, IL 60637 USA} 

\begin{abstract}
  We use a combination of high-resolution gasdynamics simulations of
  high-redshift dwarf galaxies and dissipationless simulations of a
  Milky Way sized halo to estimate the expected abundance and spatial
  distribution of the dwarf satellite galaxies that formed most of
  their stars around $z\sim 8$ and evolved only little since then.
  Such galaxies can be considered as {\it fossils\/} of the
  reionization era, and studying their properties could provide a
  direct window into the early, pre-reionization stages of galaxy
  formation.  We show that $\sim 5-15\%$ of the objects existing at
  $z\sim 8$ do indeed survive until the present in the \MW\ like
  environment without significant evolution. This implies that it is
  plausible that the fossil dwarf galaxies do exist in the Local
  Group. Because such galaxies form their stellar systems early during
  the period of active merging and accretion, they should have
  spheroidal morphology regardless of their current distance from the
  host galaxy. Their observed counterparts should therefore be
  identified among the dwarf spheroidal galaxies. We show that both
  the expected luminosity function and spatial distribution of dark
  matter halos which are likely to host fossil galaxies agree
  reasonably well with the observed distributions of the luminous
  ($L_V\ga10^6\Lsun$) Local Group fossil candidates near the
  host galaxy ($d\la200\dim{kpc}$). However, the predicted abundance is
  substantially larger (by a factor of 2-3) for fainter galaxies
  ($L_V<10^6\Lsun$) at larger distances ($d\ga300\dim{kpc}$).
  We discuss several possible explanations for this discrepancy.
\end{abstract}


\section{Introduction}
\label{sec:intro}

Abundance and spatial distribution of the faintest dwarf galaxies
provide information and constraints on the galaxy formation processes
and may give important clues to the nature of dark matter.
Cosmological simulations \citep{kkvp99a,mggl99a} and
semi-analytic models of galaxy formation
\citep{kwg93a,bkw00a,s02a,bflb02a}
predict the number of gravitationally bound, dark matter clumps (or
{\it subhalos\/}) around the \MW\ type galaxies larger, by up to an
order of magnitude, than the observed number of dwarf galaxies in the
Local Group.  In addition, the predicted spatial distribution of the
dark matter subhalos around their hosts is more extended than the
observed distribution of the Local Group satellites
\citep{tsb04b,kgk04a}, indicating that strong 
spatial bias in galaxy formation should have been present if 
the CDM scenario is correct.

Formation of galaxies in small-mass, dwarf-sized halos should be
affected by heating from the ultra-violet (UV) radiation and internal
feedback from stars. UV heating, in particular, is expected to be
highly important after the epoch of reionization when the entire
volume of the Universe is affected by the cosmic UV background. The UV
heating leads to dramatic increase of the characteristic Jeans or {\it
filtering\/} mass, $M_{\rm F}$, of intergalactic gas
\citep{gh98a,g00c}. The dark matter halos with masses $M\lesssim
M_{\rm F}$ (or circular velocities $V_{\rm c}\lesssim V_{\rm F}$)
cannot accrete new gas and therefore do not have supply of fresh gas
to fuel the continuing star formation
\citep[e.g.,][]{e92a,tw96a,qke96a,ns97a,ki00a,g00c,dhrw04a}. In
addition, the UV background can evaporate the existing gas in small
halos \citep{bl99a,sd03a,sir04a}, which would also reduce the star
formation rate.

In the presence of these galaxy formation suppression processes, the
smallest dwarf galaxies can form: (1) if their host halos have formed
sufficiently early, before the reionization epoch and UV heating
became significant \citep{bkw00a} and/or (2) if the host halos become
sufficiently massive after reionization such that UV heating effects
are not significant \citep{kgk04a}.  It is of course possible that a
halo forms early and forms stars before reionization and then stays
above filtering mass for an extended period of time after reionization
due to continuing merging and accretion. Such galaxies would exhibit
continuous, albeit perhaps episodic \citep[see][]{kgk04a}, star
formation over an extended period of time.

It is also possible that some galaxies form stars before reionization
and then not evolve much in terms of their total and stellar mass,
with their stellar populations evolving mostly passively.  Most of
such galaxies do not survive until $z=0$ but are disrupted via merging
and tidal heating, contributing to the formation of the bulge, stellar
halo, and halo population of globular clusters of the host galaxies
\citep[e.g.,][]{ws00a,bkw01a,kg05a,bj05a,rea05b,mdmzs05a}.  However, a
fraction of them can survive. \citet{rg05a} found that many of the
dwarf galaxies identified in the high-resolution gas dynamic
simulation at $z\approx 8$ do indeed resemble a subset of the observed
old dwarf spheroidal galaxies in the Local Group in their kinematic,
structural, and chemical properties.

Such galaxies would then represent {\it fossils\/} of the reionization
era, and studying their properties could provide a direct window into
the early, pre-reionization stages of galaxy formation.  As such, it
would be extremely interesting to identify and study this population
in the Local Group. In this paper, we present theoretical estimates
for the abundance and spatial distribution of the fossil galaxies
using a combination of sophisticated galaxy formation simulations at
high redshifts, which include non-equilibrium chemistry and
self-consistent, three-dimensional radiative transfer and
high-resolution dissipationless simulations of a \MW-sized halo,
which allows us to track small-mass dwarf dark matter halos from the
pre-reionization epoch ($z\sim 10$) to their present-day \MW-like
environment.

\section{Method}
\label{sec:method}

\subsection{Simulations}
\label{sec:sims}

In this paper we combine results from several simulations to connect
models of high-redshift dwarf galaxies with the present-day Local
Group. The simulation we use for modeling high-redshift dwarf galaxies 
is described in detail in \citet{rgs02a,rgs02b} as run
``256L1p3'' and was used in \citet{rg05a} as the main basis for
their conclusions. 

The simulation includes dark matter, gas, stars, the feedback of star
formation on the gas chemistry and cooling, and the
spatially-inhomogeneous and time dependent radiative transfer of
ionizing and Lyman-Werner band photons - the latter being crucial for
realistic modeling early star formation. The run followed $256^3$ dark
matter particles, an equal number of baryonic cells, and more than
700,000 stellar particles. The mass resolution of the simulation is
$900\Msun$ in baryons, and real comoving spatial resolution
(twice the Plummer softening length) is $150h^{-1}\dim{pc}$ (which
corresponds to a physical scale of $24{\rm\ pc}$ at $z=8.3$) in a
computational box with of $1h^{-1}$~Mpc on a side. Hereafter,
we refer to this simulation as L1.

Clearly, the $1\chimp$ box is too small to be a representative volume
of the universe at any moment in time. The box is appropriate for the
purposes of this study because properties of the small-mass dwarf
galaxies formed in this simulation are determined primarily by the
physics of gravitational collapse, gas cooling, star formation and
feedback and should not be affected by the box size. However, these
galaxies would not be able to reionize even such a small volume of
space. Therefore, \citet{rg05a} introduced an additional source of
ionizing radiation within the computational box, corresponding to a
star-forming galaxy with constant star formation rate of
$1\Msun/{\rm yr}$. The source was switched on at $z=9.0$,
and by $z=8.3$ the entire simulation box was completely ionized. At
that time, the computational volume contains about two dozen galaxies
with luminosities between $10^5$ and $10^7\Lsun$ and circular
velocities $\lesssim 30\dim{km/s}$. These galaxies lost their diffuse
IGM due to the UV feedback of reionization \citep{g00c}, and stopped
forming stars. \citet{rg05a} argue that these galaxies can be
identified with the population of fossil galaxies in the Local Group
dwarfs.

The specific value for the redshift at which $1\chimp$ box is reionized
($z=9$) is, of course, somewhat arbitrary, but it is important to 
note that it should {\it not\/} be
identified with the redshift of reionization of the entire
universe. It is likely that the region around the Local Group progenitor 
is reionized considerably earlier than the 
universe as a whole, because progenitors of the \MW-sized objects
can easily create their own \HII\ regions of the order of $1\chimp$ by
$z\sim9$ or even earlier\footnote{This indeed happens in a
gasdynamics simulation of a \MW-sized progenitor with radiative transfer
simulation analyzed by the authors \citep[the simulation is described 
briefly in][]{t06a}.}. 

A $1\chimp$ box size is also too small to be representative of the
Local Group environment. In the concordance LCDM
cosmology\footnote{$(\Omega_{\rm m}, \Omega_{\Lambda}, h,
\sigma_8)=(0.3, 0.7, 0.7, 0.9)$.} adopted throughout this paper, this
box contains only $1.2\times10^{11}\Msun$, whereas the mass of
the Local Group is estimated to be about $3\times10^{12}\Msun$
\citep[e.g.,][]{kzs02a}.  In order to make a connection between $z=8$
universe and the \MW-like environment at $z=0$, we use a
dissipationless (dark matter only) simulation with $20\chimp$ box
(hereafter L20), which was run using the Adaptive Refinement Tree
$N$-body code \citep[ART, ][]{kkk97a,k99a} to follow the evolution of
the \MW-sized halo with high resolution.  A Lagrangian region
corresponding to a sphere of $5R_{\rm vir}$ of a \MW-sized halo
at $z=0$ in a low-resolution simulation was resimulated with higher
resolution using multiple mass resolution technique \citep[this
simulation is Box20 run presented in][]{p05a}. In the high-resolution
region the mass of the dark matter particles is $m_{\rm p}=6.1\times
10^5h^{-1}\Msun$, corresponding to effective $1024^3$
particles in the box, at the initial redshift of the simulation
($z_{\rm i}=70$). The high mass resolution region was surrounded by
layers of particles of increasing mass with a total of four particle
species.  The simulation starts with a uniform $256^3$ grid covering
the entire computational box. Higher force resolution is achieved in
the regions corresponding to collapsing structures by recursive
refining of all such regions using an adaptive refinement
algorithm. Only regions containing highest resolution particles were
adaptively refined.  The maximum of nine refinement levels was reached
in the simulation corresponding to the peak formal spatial resolution
of $150h^{-1}$ comoving parsec.  The \MW-sized host halo has the
virial mass of $1.4\times 10^{12}h^{-1}\Msun$ (or $2.3$
million particles within the virial radius) and virial radius of
$230h^{-1}\dim{kpc}$ at $z=0$.

\subsection{Simulation and Data Samples}
\label{sec:samples}

Before we compare results of the cosmological simulation with
observational data, we consider the limits of its numerical
resolution.  In this paper we choose to characterize dark matter halos
by their maximum circular velocity $\vm$, because it is well-defined
while definition of the halo mass is rather arbitrary.  For the
low-mass halos that may host dwarf galaxies, the maximum circular
velocity function (the number of halos per unit $\vm$) is expected to
be close to a power-law \citep[e.g.,][]{kkvp99a}. Thus, strong
deviation from a power-law at small halo masses indicates resolution
limit of the simulation \citep[see, e.g.,][]{rea05a}.

\begin{figure}[t]
\plotone{\figname{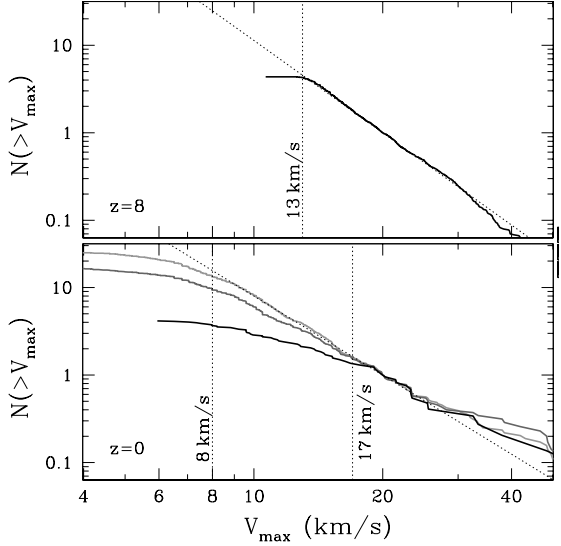}}
\caption{Cumulative maximum circular velocity functions for the L20
run at $z=8$ (top panel) and $z=0$ (bottom panel). Velocity functions
at $z=0$ are shown for three subsamples: halos within $1\dim{Mpc}$ of
the parent halo ({\it light gray line}), $300\dim{kpc}$ ({\it dark
gray line}), and $100\dim{kpc}$ ({\it black line}). Tilted {\it thin
dotted lines} show power-law $N(>V_{\rm max})\propto V_{\rm max}^3$,
while the {\it vertical dotted lines} mark characteristic values of
$\vm$ corresponding to the resolution limit of the simulation, 
as discussed in the text. The velocity functions in both panels
are normalized to unity at $\vm=20\kms$.  }
\label{figNV}
\end{figure}

Figure \ref{figNV} illustrates this point. It shows the cumulative
maximum velocity function (i.e., the number of dark matter halos with
$\vm$ above a given value) at $z=8$ and $z=0$ for the L20
run. The strong deviation from the power-law at $z=8$ can be seen at
$\vm\lesssim 13\kms$, indicating the resolution limit of the L20
simulation at this epoch. At $z=0$, the issue is more complicated
because resolution depends on the distance to the \MW-sized halos halo.
For example, an isolated halo will exist even if its internal
structure is poorly resolved, but it may be prematurely disrupted when
it falls into a parent halo and experiences tidal heating.  Three
lines in the lower panel of Fig.\ \ref{figNV} do indeed show that
while the L20 simulation is sufficient to resolve dark matter halo
down to about $8\kms$ within $1\dim{Mpc}$ of the parent
halo\footnote{The improvement in resoltion limit, from $13\kms$ at
$z=8$ to $8\kms$ at $z=0$, is likely due to the fact that in a
adaptive mesh refinement simulation we use, the resolution becomes
somewhat better with time.}, it is complete only down to $17\kms$ for
halos within $100\dim{kpc}$ of the \MW-sized halo.

We can simply correct for the incompleteness by a velocity dependent
factor that is required to restore the power-law form of $N(>\vm)$
function (in other words, the correction factor is the ratio of the
power-law and the actual velocity function measured in
simulations). To illustrate the effect of incompleteness we will use both
the uncorrected and the corrected samples, which should bracket the
correct result. As we show below, our results are not highly sensitive 
to this correction even for the $d<100\kms$ sample, for which the correction
is the largest. Note also that the correction only becomes large at
$\vm\lesssim 15\ {\kms}$, while none of the observed dwarf galaxies in 
the Local Group are likely to have $\vm$
significantly below $15\kms$ \citep[e.g.,][]{swts02a,zb03a,kea04a,mdmzs05a}

\begin{figure}[t]
\plotone{\figname{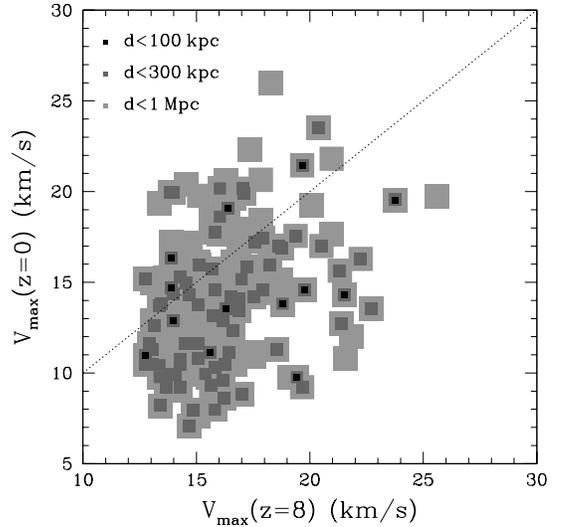}}
\caption{Distribution of maximum circular velocities for dark matter
halos from run L20 at $z=8$ versus $z=0$. Halos at $z=0$ are
distinguished by their distance to the parent halo: halos within
$1\dim{Mpc}$ ({\it large light gray squares}), within $300\dim{kpc}$
({\it intermediate darker gray squares}), and within $100\dim{kpc}$
({\it small black symbols}). Only halos with $\vm(z=8)>13\kms$
are shown. }
\label{figVV}
\end{figure}
Figure \ref{figVV} shows the sample of dark matter halos from the
L20 run we use in this paper. Three shades of gray show three
subsamples according to the distance from the parent halo at $z=0$. As we
mentioned above, we only select halos that have $\vm>13\kms$ at $z=8$,
so our sample is complete at high redshift. For these halos, $\vm$ at
$z=0$ reaches down to $\vm\approx8\kms$, so our sample of halos with
$d<1\dim{Mpc}$ is effectively complete, while subsamples with smaller
limits on $d$ are incomplete to various degrees.

\begin{figure}[t]
\plotone{\figname{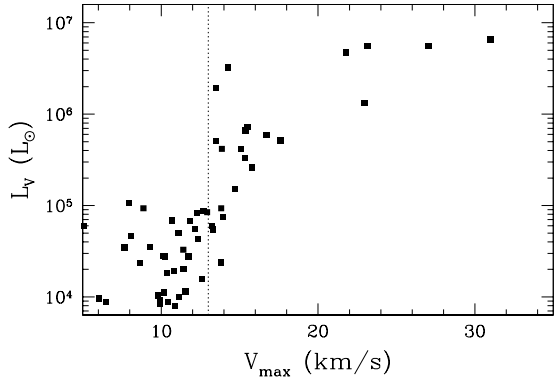}}
\caption{The luminosity - maximum circular velocity relation for the
  dwarf galaxies in the run L1 at $z=8$. The {\it thin dotted line\/} marks
  the resolution limit of our sample at $\vm=13\kms$ (see text for details).
}
\label{figVL}
\end{figure}

Finally, the L1 simulation allows us to relate the stellar
luminosities of model dwarf galaxies to the maximum circular
velocities of their dark matter halos at $z=8$. The latter are
measured with the same halo finder algorithm used to identify
halos in the L20
run. Figure \ref{figVL} shows the distribution of these two properties
for the model dwarf galaxies from the L1 run. We marked the
completeness limit of the L20 run at $z=8$ ($\vm=13\kms$) with the
thin dotted line, and throughout this paper we only use dwarf galaxies
above this limit. Fortunately, all model dwarfs with
$L_V>10^5\Lsun$ fall within this range.

The objects shown in Figures\ \ref{figVV} and \ref{figVL}, thus,
constitute the sample of the simulated galaxies that we use in this
paper. Our main task now is to compare these model galaxies to the
observed dwarf galaxies in the Local Group that \citet{rg05a}
identified as reionization fossils. The observed sample includes
Draco, Phoenix, Sculptor, Sextans, Tucana, and Ursa Minor for the
\MW\ subgroup, and And I-III,V,VI,IX, Antlia, Cassiopea, Cetus,
EGB0427, and SagDIG  
for the Andromeda subgroup. These galaxies are selected because they
are dim (V-band luminosity below $10^7\Lsun$) and because they
formed at least 70\% of their stars in one single burst at an early
epoch (which is identified with the reionization epoch). These
galaxies therefore represent reasonable candidates for the
reionization fossils. This identification is, of course, somewhat
uncertain because the absolute ages of $10-12\dim{Gyr}$ old stars are
not measured with sufficient accuracy. Nevertheless, we will assume in
the remainder of this paper that the observational samples are not
highly contaminated by non-fossil galaxies. We refer reader to
\citet{rg05a} for more details on how the observational sample is
selected.

\subsection{Survival Probability}
\label{sec:suvpro}

In order to test the fossil scenario, we need first to measure the
probability for a low-mass dark matter halo to remain relatively
intact (i.e., not to evolve significantly) from $z\sim8$ until the
present time, which \citet{rg05a} called the ``survival
probability.'' We use the L20 simulation to measure the survival
probabilities for dark matter halos. Since in the dissipationless
simulation we cannot identify halos with luminous galaxies, we can
only measure the survival probability as a function of the maximum
circular velocity. We will use the mapping between circular velocities
and galaxy luminosities provided by the L1 run (see
Figure~\ref{figVL}) to calculate from $\vm$ to luminosity.

We measure the survival probability by tracking evolution of the
individual halos from $z=8$ to the present time\footnote{In the
next section we discuss the sensitivity of our results to the
specific choice of the redshift at which fossils are identified.}
using the method developed by \citet{kgk04a}. To track the halos we
use 156 simulation outputs equally spaced in expansion factor from
$z=14$ to $z=0$.

The tracks then can be used to estimate the fraction of halos existing
at $z=8$ which remain distinct, bound objects at $z=0$. For detailed
comparison with observations, it is useful also to differentiate halos
with respect to the distance to the \MW-sized progenitor, as significant
morphological segregation is known to exist in the Local Group. Thus,
we define the survival probability of halos within a given distance
$d$ (at $z=0$) from the parent halo as
\begin{equation}
  \ps(d,\vm) = \frac{N_{\rm H}(8\rightarrow0,\vm)}{N_{\rm H}(z=8,\vm)},
  \label{psdef}
\end{equation}
where $N_{\rm H}(8\rightarrow0,\vm)$ is the number of halos that exist
both at $z=0$ and $z=8$ within  distance $d$ (measured at $z=0$)
from the parent halo as a function of $\vm$, and $N_{\rm H}(z=8,\vm)$
is the total number of halos at $z=8$ located in the region of
space (the ``Lagrangian region'') that ends up as a sphere
of radius $d$ centered on a parent halo at $z=0$. 

It is important to stress that in equation (\ref{psdef}) the distance
$d$ is measured at $z=0$, while the maximum circular velocity $\vm$ is
measured at $z=8$.  Operationally, in order to find all halos within
the $z=8$ Lagrangian region of sphere of radius $d$ around the
\MW-sized halo at $z=0$, we tag all dark matter particles within the
sphere and identify all dark matter halos at $z=8$ located in the
region of space occupied by the tagged dark matter particles.

Note that in this study we calculate and apply the survival probability
to the small-mass dwarf objects of $\vm\lesssim 30\kms$. Such 
galaxies are dark matter dominated and effects of baryon dissipation 
on the inner mass distribution, $\vm$, and on the survival
of halos at small distances to the host \citep[e.g.,][]{mbsd05a} should be
small. 

\begin{figure}[t]
\plotone{\figname{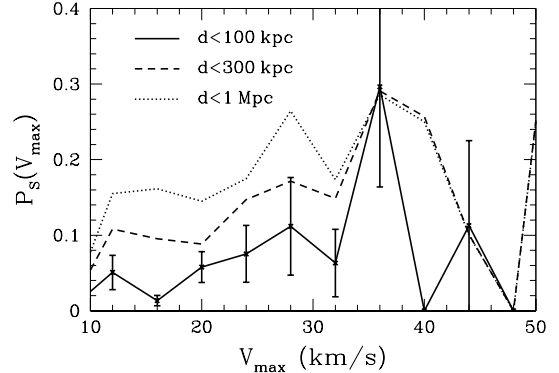}}
\caption{The survival probability, $\ps(\vm)$, of {\it all dark matter
  halos\/} that end up within $100\dim{kpc}$ (solid line),
  $300\dim{kpc}$ (dashed line), and $1\dim{Mpc}$ (dotted line) from
  the parent halo in the L20 run. The error bars are Poisson
  only. (Reminder: the maximum circular velocity $\vm$ is measured at
    $z=8$.)}
\label{figPA}
\end{figure}

As an example, Figure \ref{figPA} shows the survival probability $\ps$
as a function of $\vm(z=8)$ for all dark matter halos from the L20
run.  The survival probability ranges from a few to $\approx 25$ per
cent and there is a clear trend with the distance to the parent
halo. Our measurement of $\ps(\vm)$ becomes highly noisy for
$\vm>35\kms$ due to small numbers of halos.

\subsection{Defining a Fossil}
\label{sec:fossil}

The survival probability is of course defined for all dark matter
halos, not only for those that host fossil galaxies. In order to
compute the survival probability for fossils only, we need an
operational definition of a reionization fossil applicable to
dissipationless simulations. We consider a dwarf galaxy to be a fossil
if it experiences little or no star formation after reionization. This
would be the case if the galaxy does not accrete fresh gas to fuel its
continuing growth and star formation. The accretion of gas on dark
matter halos is controlled by the filtering mass \citep{g00c}, which
is easily calculable for a representative volume of the universe, but
is not well defined for a region that collapses into a \MW-sized
galaxy. As we noted above, such region can be reionized considerably
earlier. Nevertheless, after the region is ionized, the characteristic
mass and circular velocity below which accretion is suppressed should
be in the range of $\sim 30-50\ \kms$, as was shown by many studies
\citep{tw96a,qke96a,ns97a,ki00a,g00c,dhrw04a}.  Therefore, we define a
reionization fossil as a halo that never increased its maximum circular
velocity above a given fixed threshold $\vf$ (note that we use $\vf$
here to distinguish the threshold circular velocity for fossils from
the filtering velocity, $\vtf$),
\begin{equation}
  \max[\vm(z)] < \vf.
  \label{tfdef}
\end{equation}
This definition can be easily implemented for the dissipationless
simulations we use, as we track the evolution of $\vm$ for all halos.

\begin{figure}[t]
\plotone{\figname{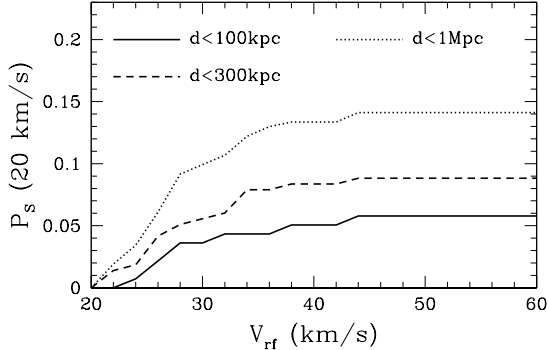}}
\caption{The survival probability $\ps$ as a function of the threshold
  parameter $\vf$ ({\it gray line}) for halos with
  $\vm=20\pm2\dim{km/s}$ from the L20 run. Three lines correspond to
  the three halo subsamples: halos within $1\dim{Mpc}$ ({\it dotted
  line}), $300\dim{kpc}$ ({\it dashed line}), and $100\dim{kpc}$ of
  the parent halo ({\it solid line}), respectively.  }
\label{figCV}
\end{figure}

Figure \ref{figCV} shows the dependence of the survival probability on
this parameter for halos with $\vm=20\pm2\dim{km/s}$.  In the limit of
large $\vf$, for example, 14\% of all halos present at $z=8$ survive
as separate object by $z=0$ for the $d<1\dim{Mpc}$ sample, while the
rest grow into larger objects or get disrupted before $z=0$. For
objects closer to the parent halo the survival probability decreases,
since merging and disruption are expected to be more efficient in the
higher density environment.

The survival probability levels off for $\vf>50\kms$, because if the
value of the threshold $\vf$ is set too high above the value of $\vm$
for the halos ($20\kms$ for Fig.\ \ref{figCV}), it becomes
irrelevant. Note also that the survival probability is a weak function
of the filtering velocity for $\vf\gtrsim 30\,\kms$.  Our results
thus are not particularly sensitive to the specific choice of $\vf$
as long as it is in this range. We demonstrate this explicitly below
(see Fig.\ \ref{figLZ}).  In what follows, we adopt $\vf=30\kms$ as
our fiducial value, as it is sufficiently small to ensure that halos
that we identify as fossils have minimal accretion of fresh gas after
reionization.

\begin{figure}[t]
\plotone{\figname{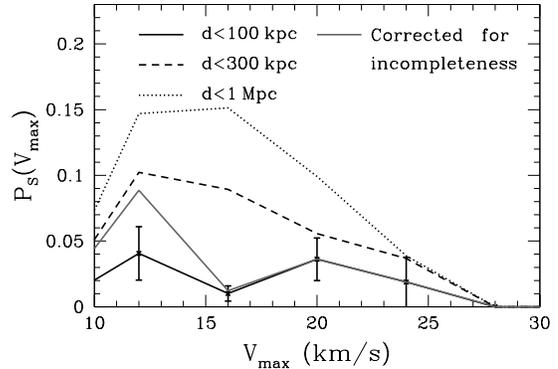}}
\caption{The same as in Fig.\ \protect{\ref{figPA}}, but for the reionization
  fossils according to the definition (\protect{\ref{tfdef}}). The
  {\it solid gray line} shows the survival probability for the
  $d<100\dim{kpc}$ when the correction for incompleteness
  is taken into account.}
\label{figPF}
\end{figure}

Figure \ref{figPF} shows the survival probability as a function of
$\vm(z=8)$ (similar to Fig.\ \ref{figPA}) but for the reionization
fossils defined using equation~(\ref{tfdef}) with $\vf=30\kms$. Note
that $\ps$ approaches zero for $\vm$ approaching the value of the
threshold $\vf=30\kms$, simply because the definition does not allow
existence of such fossils.  At $\vm\la15\kms$ the incompleteness of
the $d<100\dim{kpc}$ sample becomes non-negligible, and the survival
probability at lower values of the maximum circular velocity becomes
uncertain by up to a factor of two. This uncertainty does not
significantly affect our results since most of our galaxy sample has
$\vm\ga15\kms$.

 The survival probability ranges from a few per cent for objects
located close to the \MW\ host to $\approx 15\%$ for more distant
halos. Thus only a small fraction of objects at $z=8$ would survive to
be true fossils at the present epoch.  The survival probability
distribution shown in Figure~\ref{figPF} is the main ingredient of our
model for the luminosity function and spatial distribution of fossil
galaxies presented in the next section.

\section{Results}
\label{sec:results}

Given the survival probability of fossils as a function of $\vm$ and
$d$ from the L20 run, we can use the galaxy formation run L1 to
compute the luminosity function of fossils as a function of distance
from the parent galaxy. Specifically, we use the distribution of
high-redshift dwarfs in the $L_V$ - $\vm$ plane shown in Fig.\
\ref{figVL} to compute the cumulative luminosity function of fossils
at $z=0$ within a given Lagrangian region with mass $M_{\rm LR}$:
$$
  N_{\rm LR}(>L_V) = \left(\frac{M_{\rm
      LR}}{1.2\times10^{11}\Msun}
  \right)
  \int_{L_V}^{10^7\Lsun} dL
$$
\begin{equation}
  \phantom{WW}\int d\vm \ps(\vm) 
  \frac{d^2N_8(L,\vm)}{dLd\vm},
  \label{nleq}
\end{equation}
where $1.2\times10^{11}\Msun$ is the mass of the computational
volume of the L1 run and $d^2N_8(L,\vm)/dL/d\vm$ is the luminosity
function of high redshift dwarfs at $z=8$ in the L1 run per unit
$\vm$. For the small number of objects we have in the L1 run, 
$$
  \frac{d^2N_8(L,\vm)}{dLd\vm} = \sum_j \delta(\vm-\vm^j)\delta(L_V-L_V^j),
$$ 
where $\vm^j$ and $L_V^j$ are the maximum circular velocity and the
V-band luminosity of a dwarf galaxy $j$.

The upper limit of $10^7\Lsun$ in the integral over the
luminosity is motivated by \citet{rg05a}, who found that no fossils
exist at higher luminosities in the Local Group. The limits of the
integral over $\vm$ are formally over all existing halos, but the
survival probability is non-zero only over a finite range of
$13\kms<\vm<\vf$ due to our definition of the reionization fossil and
our completeness limit.

In order to define the amplitude of the cumulative luminosity function
for the Local Group, we need to know the mass of the Lagrangian region
that ends up within a given distance from the parent galaxy (the \MW\
or Andromeda). To estimate this mass, we use the dynamical models
of the \MW\ and Andromeda from \citet{kzs02a}. For their
preferred models (A$_1$ or B$_1$ for the \MW\ and C$_1$ for Andromeda), we
find the total mass with $100\dim{kpc}$ from the \MW\ or
Andromeda (summed over these two galaxies) to be
$1.4\times10^{12}\Msun$, and the total mass within
$300\dim{kpc}$ to be $2.7\times10^{12}\Msun$. For larger
distances, extrapolation is required. In the L20 run, about 50\% more
mass is between $300\dim{kpc}$ and $1\dim{Mpc}$ around the parent halo
compared to the total mass within $300\dim{kpc}$. This number is also
consistent with the simple extrapolation of NFW profiles around the
\MW\ and Andromeda. Thus, if the two galaxies were isolated, the
total mass of the Local Group within $1\dim{Mpc}$ would be $\approx
4.0\times10^{12}\Msun$. However, since the distance between
the \MW\ and Andromeda is about $780\dim{kpc}$, there is a
substantial overlap between the two spheres of $1\dim{Mpc}$ in radius
centered on the two galaxies, and most of the mass of both galaxies is
within the overlap region. Although an accurate estimate is difficult,
we adopt the mass $3.0\times10^{12}\Msun$ as a reasonable
estimate of the total mass of the two galaxies within $1\dim{Mpc}$
from each of them.

We first check the sensitivity of our results to adopted values of
free parameters: the starting redshift $z_s$ at which we identify the
future fossils and the fossil threshold $\vf$. Indeed, although $z=8$
is a reasonable choice for characteristic pre-reionization epoch,
there is nothing special about it. This specific value is simply the
epoch when the L1 run was stopped in \citet{rg05a} for physical and
computational reasons. In addition, depending on the particular
evolutionary histories of the real \MW\ and Andromeda, the Local
Group region could have been reionized at different redshifts.

\begin{figure}[t]
\plotone{\figname{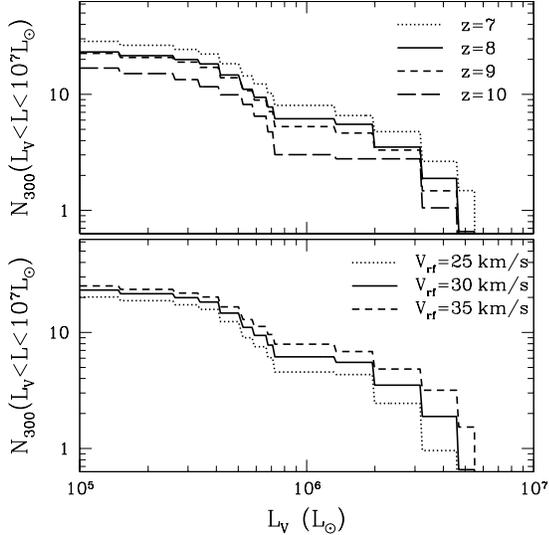}}
\caption{The cumulative luminosity functions of fossil dwarfs located
  within $300\dim{kpc}$ from the parent halo at $z=0$ as a function of
  the starting redshift ({\it top panel}) and the fossil threshold
  parameter $\vf$ ({\it bottom panel}).  }
\label{figLZ}
\end{figure}

The top panel of Figure \ref{figLZ} shows the cumulative luminosity
functions for fossils located within $300\dim{kpc}$ of the parent halo
at $z=0$, calculated using survival probabilities derived from the L20
run. Four lines correspond to the four values of starting redshift:
$z_s=7$, 8, 9, and 10. Namely, the solid line in Fig.\ \ref{figLZ} is
obtained using the equations above, while the dotted line ($z=7$) is
obtained from the same equations but with $z=8$ replaced by $z=7$
everywhere. The figure shows that the dependence on the specific value
for the starting redshift for measuring the survival probabilities is
quite weak for $z_s\la9$. In making this figure, we used the same
$L_V-\vm$ distribution from Fig.\ \ref{figVL}, since we cannot rerun
the L1 simulation for different reionization redshifts (due to its
expense). However, a preliminary analysis of different galaxy
formation simulation (Tassis et al. 2006, in preparation) indicates
that redshift dependence of the $L-\vm$ relation for high-$z$ dwarfs is
given by $L\propto \vm^3(1+z)^{-1.5}$. Depending on how different the redshift
of reionization is from our fiducial value of $z=8$, the difference in
the relation can be approximated by this redshift scaling.  Thus, for
example, if we change redshift from $z=8$ to $z=10$ the difference in
luminosity for a given $\vm$ is only $35\%$. If however, the change is
to $z=15$ the difference is a factor of two.  This uncertainty can
therefore be assumed as uncertainty of a factor of two in
the luminosity we assign to the fossil dwarfs.

\begin{figure}[t]
\plotone{\figname{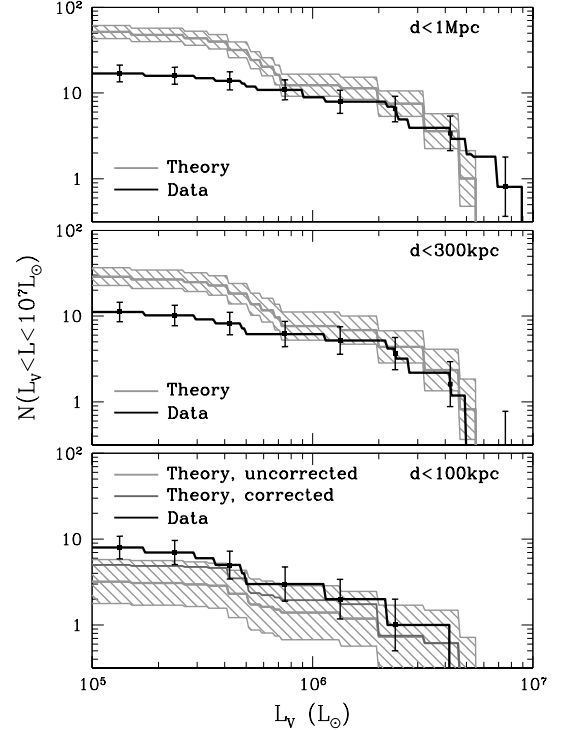}}
\vspace{-5mm}
\caption{The cumulative luminosity functions of fossil dwarfs located
  within $1\dim{Mpc}$ ({\it top panel}), $300\dim{kpc}$ ({\it middle
  panel}), and $100\dim{kpc}$ ({\it bottom panel}) from the parent
  galaxy at $z=0$. The {\it black lines} with error-bars are the
  observational data for the Local Group dwarfs, while the {\it light gray
  bands} show model predictions with $1\sigma$ error-bars,
  uncorrected for incompleteness. The {\it dark gray line} in the bottom
  panel shows the model prediction corrected for the incompleteness.  }
\label{figLF}
\end{figure}

At higher redshifts the dependence is much stronger. We can of course
expect that as we increase the starting redshift to $z_s\sim 10-20$
the number of halos that do not change appreciably to the present
epoch becomes very small. Our choice for the lower value of $z_s$ is
fairly straightforward. The universe is known to be fully reionized at
$z\approx6$, so $z=7$ is quite close to the lowest possible
value. Indeed, it is highly unlikely that the vicinity of the Local
Group should be one of the last patches of neutral gas in the universe
to be re-ionized at $z<7$. The upper limit of $z\sim9$ is less
constrained, but is generally a reasonable number to expect a region
of about $1\chimp$ across to be reionized by the UV radiation from a
\MW\ progenitor. Of course, other sources of radiation may cause even
earlier partial or complete ionization of the Lagrangian region around
the \MW\ or Andromeda. The predicted abundance of fossils would
be correspondingly smaller, as we discuss in the next section.
The bottom panel of Figure~\ref{figLZ} shows dependence of the 
luminosity function of fossils on the adopted value of $\vf$. The
dependence is weak and the amplitude of the luminosity function 
changes by less than a factor of two for $\vf$ from $25\,\kms$ to 
$35\,\kms$.

\begin{figure}[t]
\plotone{\figname{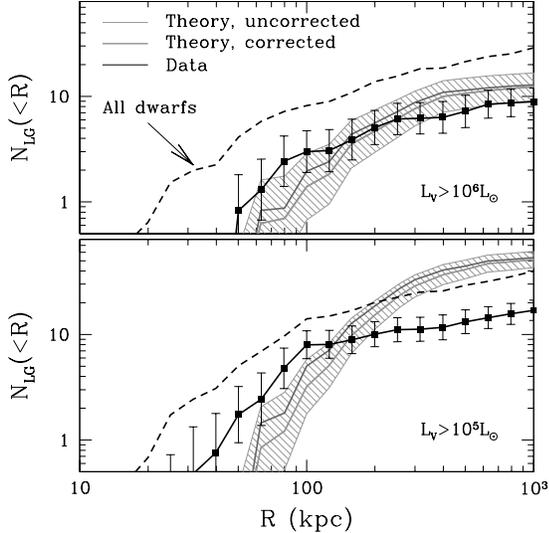}}
\caption{The radial distribution of reionization fossils with
  $L_V>10^6\Lsun$ (top panel) and $L_V>10^5\Lsun$
  (bottom panel). The light gray bands and dark gray lines show theoretical
  expectations without and with correction for incompleteness. Black
  lines with error-bars show the observational data on Local Group
  dwarfs identified as reionization fossils by
  \protect{\citet{rg05a}}, while the black dashed line show all
  (whether a fossil or not) Local Group dwarfs.  
}
\label{figRD}
\end{figure}

Figure \ref{figLF} presents the main result of this paper: the
cumulative luminosity function of reionization fossils in the Local
Group within three different distances from the parent galaxy. The
observational data for the Local Group dwarfs are taken from
\citet{m98a} and \citet{mi05a}, as described in
\S~\ref{sec:samples}. Since observational uncertainties on the
distances (from the \MW) to some of Andromeda satellites are
comparable to their distance to Andromeda, there is an uncertainty on
whether a given dwarf galaxy is within, say, $100\dim{kpc}$ from
Andromeda. To take this uncertainty into account, we construct 100
random realizations of the Local Group in which dwarf spheroidal
galaxies are distributed over the distance (from the \MW) with a
log-normal distribution of the width corresponding to the distance
measurement error.  

The errorbars that we assign to the observed luminosity function
include both the Poisson fluctuations and the error measured from the
dispersion between 100 random realizations of galaxy distances.  The
predictions of our model for the fossil galaxies is shown by the
hatched band and includes Poisson errors in the number of galaxies
from the L1 run and Poisson errors in the survival probability due to
the limited number of halos in the L20 run. Since we have only one
dissipationless simulation, additional uncertainty due to cosmic
variance should exist as well. However, we expect it to be smaller or
comparable to the Poisson errorbars shown here.  The dark gray line in
the bottom panel of Fig.\ \ref{figLF} shows the mean value of the
theoretical prediction with the incompleteness correction included.

We can draw two straightforward conclusions from Figure~\ref{figLF}.
\begin{enumerate}
\item The expected number of fossils with $L_V>10^6\Lsun$ is
in reasonable agreement with the observed luminosity function of the
Local Group dwarfs, which are likely to be fossils, for all distances
from the parent galaxy.
\item The predicted number of fossils with
$10^5\Lsun<L_V<10^6\Lsun$ is also in reasonable
agreement with observations within $100\dim{kpc}$, but is larger by a
factor 2 to 4 for $d<300\dim{kpc}$ and $d<1\dim{Mpc}$.
\end{enumerate}
These conclusions are not sensitive to our incompleteness correction, as
it is smaller than our estimated error bars.

Figure \ref{figRD} compares the predicted radial distribution of the
fossil galaxies for two luminosity ranges ($L_V>10^6\Lsun$ and
$L_V>10^5\Lsun$) to the distribution of observed Local Group
dwarfs. It is interesting to note that the distribution of observed
galaxies identified by \citet{rg05a} as fossils is quite similar to
the radial distribution of all dwarf galaxies. There is thus no
particular spatial bias exhibited by such galaxies. It is not clear
whether a strong spatial bias should exist. On the one hand, the early
forming high-redshift galaxies should be spatially biased with respect
to their host. On the other hand, the galaxies located close to the
host would be accreted early and thus may have a lower chance to
survive until the present.

The figure shows that the predicted radial distribution for the
brighter fossils ($L_V>10^6\Lsun$) agrees with observations.
For lower luminosities, however, the predicted and observed radial
distributions are quite different. Although the number of predicted
fossils at $d<100\dim{kpc}$ is similar to the observed abundance, the
observed dwarfs show a somewhat more radially concentrated
distribution. The difference is not large however and is not
statistically significant. At larger distances, on the other hand, the
model predicts far more (a factor of 3-4) fossils with
$L_V>10^5\Lsun$ than is observed. There is thus a marked
deficit of the observed galaxies at the faintest luminosities at
distances larger than $100\dim{kpc}$. These conclusions are the same if we
consider radial distribution for Andromeda or the \MW\ only, although
the statistics is poorer in this case. We discuss possible explanations
for the discrepancy in the next section. 

\section{Discussion and Conclusions}
\label{sec:discussion}

In this paper we estimate the expected abundance and spatial
distribution of the reionization fossils: dwarf galaxies that formed
most of their stars around $z\sim 8$ and evolved only little since
then. We show that $\sim 5-15\%$ of the objects existing at $z\sim 8$
do indeed survive until the present in the \MW\ like environment
without significant evolution. Because such galaxies form their
stellar systems early during the period of active merging and
accretion, they should have spheroidal morphology and their observed
counterparts are most likely to be identified among the dwarf
spheroidal galaxies. The spheroidal morphology of the fossil galaxies
is thus not related to the tidal heating by the nearby massive
galaxies and they can be found even at distances as large as a
megaparsec from their host. This could possibly explain existence of
distant dwarf spheroidal galaxies in the Local Group, such as Cetus,
Tucana, and KKR25.

We show that both the expected luminosity function and spatial
distribution of dark matter halos which are likely to host fossil
galaxies agrees reasonably well with the observed distribution of the
luminous ($L_V\ga10^6\Lsun$) Local Group fossil candidates
near the host ($d\la200\dim{kpc}$). The predicted abundance is
substantially larger (by a factor of 2-3) for fainter galaxies
($L_V<10^6\Lsun$) at larger distances
($d\ga300\dim{kpc}$). This discrepancy can have at least three
plausible explanations. We list them and the possible tests below.

\begin{description}
\item[Observations are wrong.] Observational incompleteness for faint,
  $L_V<10^6\Lsun$, fossils may potentially account for the
  difference we find, if it can be as high as a factor of 3 at
  $d>300\dim{kpc}$. \citet{wgdr04a} found that a factor of 3
  incompleteness is possible, although that number was on the very
  edge of their estimates. The same incompleteness factor would have
  to apply to the distribution of Andromeda satellites, which seems to
  be less likely. Future observational searches (for example, from
  SDSS) for dwarf spheroidals between $200\dim{kpc}$ and $1\dim{Mpc}$
  from the \MW\ {\it or\/} Andromeda, or observations of nearby
  galaxy groups with the next generation large telescopes should be
  able to test this possibility.
\item[Theory is wrong.] The abundance of fossils galaxies may be overestimated
by our model. This is possible for example if reionization occurs much
earlier than we assumed: for example, at $z>10$, as may be
  suggested by the first year WMAP data \citep{ksbb03a}. In this case, 
  the number of small-mass objects that are able to form stars before 
reionization should be much smaller. However, if
  most of dwarf galaxies that \citet{rg05a} identified as reionization
  fossils are confirmed to be such by future observations, then this work
  will places a constraint of $z\la10$ on the reionization history of
  the region of the universe that collapses into the Local Group.
\item[Spatial bias.] If we take the observational data as complete and
  at face value, they indicate that lowest luminosity fossils
  ($L_V\la10^6\Lsun$) are located preferentially near
  the parent galaxy. At first glance, such a conclusion seems
  counterintuitive, since the feedback (both radiative and kinetic)
  from the progenitor of the main galaxy is expected to suppress star
  formation in nearby galaxies. However, galaxies are also biased,
  more so at high redshift, so dwarf galaxies located closer to the
  progenitor of a large spiral galaxy will start forming sooner than
  similar galaxies in more distant regions of space. Which of the two
  opposing effects (feedback vs bias) wins is unclear, but should be
  tackled with the future ultra high-resolution cosmological simulations. 
\end{description}

The results of this study indicate that existence of the fossil
galaxies in the Local Group, in which most of the stars formed during
the earliest stages of galaxy formation, is plausible.  The abundance
of fossils predicted by our model would constitute as much as
$20-40\%$ of all the luminous Local Group dwarfs.  The improvement in
resolution of cosmological simulations should improve our estimates.
Observationally, identification and studies of the fossil galaxies can
provide an exciting opportunity of studying the properties of star
formation in primeval galaxies by direct observations in our own
cosmological backyard.

\acknowledgements 

We would like to thank Anatoly Klypin for providing us the dark matter
simulation of the \MW-sized halo used in this study.  We are grateful
to Oleg Gnedin, James Bullock, Brant Robertson, and Ben Moore for the
comments on the draft of this paper. This work was supported in part
by the DOE and the NASA grant NAG 5-10842 at Fermilab, by the HST
Theory grant HST-AR-10283.01, by the NSF grants AST-0206216,
AST-0239759, and AST-0507596, and by the Kavli Institute for
Cosmological Physics at the University of Chicago.  Supercomputer
simulations were run on the IBM P690 array at the National Center for
Supercomputing Applications (under grant AST-020018N) and the
Sanssouci computing cluster at the Astrophysikalisches Institut
Potsdam.  We are grateful to the King Kamehameha Beach Hotel in Kona
for hospitality during the initial stages of this project.  This work
made extensive use of the NASA Astrophysics Data System and {\tt
  arXiv.org} preprint server.

\bibliographystyle{apj}
\bibliography{ms}

\end{document}